# Network-on-Chip with load balancing based on interleave of flits technique

Marcelo Daniel Berejuck
Electrical Engineering Department
University of Southern Santa Catarina - UNISUL
Florianópolis-SC, Brazil
marcelo.berejuck@ieee.org

*Abstract* — **This paper presents the evaluation of a Network-on-Chip (NoC) that offers load balancing for Systems-on-Chip (SoCs) dedicated for multimedia applications that require high traffic of variable bitrate communication. The NoC is based on a technique that allows the interleaving of flits from different flows in the same communication channel, and keep the load balancing without a centralized control in the network. For this purpose, all flits in the network received extra bits, such that every flit carries routing information. The routers use this extra information to perform arbitration and schedule the flits to the corresponding output ports. Analytic comparisons and experimental data show that the approach adopted in the network keeps average latency lower for variable bitrate flows than a network based on resource reservation when both networks are working over 80% of offered load.**

*Network-on-Chip; System-on-Chip; Load balancing; Multimedia*

## I. Introduction

Deploying real-time applications on modern System-on-Chip (SoC) mandates that the designer ensure performance and temporal guarantees. Note that modern SoC platforms typically consist of multiple processors, and a communication interconnect between them. With the need to integrate an increasing variety of processing units, communication management becomes critical when an interconnect fabric must support highly diversified functions with varying latency and bandwidth requirements. Network-on-Chip (NoC) is one solution of interconnect fabric adopted by silicon industry to connect dozens of processing units heterogeneous resources [3].

NoCs can be designed to provide communication services with guarantees on throughput and latency for flows in the network, or offering best-effort communication service with no guarantees on latency and bandwidth for them. Flows related to real-time tasks usually receive such services with guarantees on throughput and latency. The prevailing design strategy to produce interconnects with guarantees for such SoCs relies on mapping the requirements of the communication flows, related to real-time tasks, to available network resources at early design stages.

On the other hand, multimedia flows are treated as Best-Effort (BE) flows. Best-effort services usually are designed focus on improvement of average latency. Hence, BE flows are prone to network congestion what has a negative effect on network load balancing, and on its performance, as well [5]. BE flows experience the resulting performance degradation as an increase of latency and loss of bandwidth. Thus, networks with BE service should have a load balancing strategy to avoid congestion.

Most of the techniques adopted so far for best-effort flows, and as a consequence for multimedia flows, usually rely on the global knowledge of the network state. Some authors state that without global knowledge of the network state, such a strategy can never assert that the network does not reach a congested state [5], and hence, can be impossible to ensure load balancing in the network. However, the knowledge of the network state cannot be possible for SoC based on NoC that works with variable bitrate in its communication channels.

Nevertheless, after over a decade developing industrial Telecom and multimedia projects, we realized that many applications in this domain work with variable bitrate in the communication channels. They would profit better from a NoC that could optimize the load balancing. During the last decade, few works addressed load balancing on Network-on-Chip. The strategy adopted by all of them were based on getting information about the network state at run-time, to deal with the load balancing.

This paper introduces a different strategy to deal with the load balancing for SoCs based on Network-on-Chip with high traffic of variable bitrate in the communication channels. It is based on the hypothesis that the load balancing could be improved at design-time for those SoCs, even without checking the network state. The focus was 2D mesh networks. The evaluated NoC architecture, called RTSNoC [18], is based on the interleaving of flits from different flows in the same communication channel between routers. To identify those flits, extra bits were added for all of them, such that every flit carries routing information. The evaluation of the RTSNoC demonstrates that the average latency for best-effort flows with variable bitrate in the communication channels of the network is improved when the network is under high traffic. It also demonstrates that the RTSNoC latency can be consistently predicted for each flow in the network without requiring resource reservation.

The remainder of this paper is organized as follows: in Section II we outline the strategies for 2D mesh networks based on information about the network state at run-time adopted by other authors, and that were introduced so far. Section III presents RTSNoC architecture and introduces its components. Section III introduces the concept of interleaving of flits and a definition for latency in NoCs. Section IV-B introduces the analytic expression of worst-case latency (WCL) on RTSNoC. Section V introduces the experimental results, and Section VI closes the paper with our conclusion.

## II. RELATED WORK

This Section briefly introduces the strategies based on the measurement of the network state at run-time, proposed by other authors. We understand that this overview will help to differentiate between these strategies and the design-time strategy we have adopted, and will be introduced in Section III.

The first work we are introducing in this Section belongs to the authors [1]. They proposed a strategy for load balancing in NoCs based on link utilization. The strategy is based on communication service level called Congestion-Controlled Best-Effort (CCBE) that allows control of offered load based on critical shared resource utilization measurements. They use the link utilization as a congestion measure. Such measurements are performed by hardware probes and are carryout to a controller by guaranteed service connections in the NoC to assure that this communication is not subject to congestion. The path from the controller to the processing unit to communicate the computed loads can be implemented in a similar way. The controller, a Model Predictive Controller (MPC), determines the appropriate loads for the CCBE connections. The method introduced by those authors requires that routing in the NoC cannot be dynamic.

The authors [2] introduced a flow control scheme for best-effort traffic in NoC based on source rate utility maximization. They did a model of the flow control as a utility-based maximization problem, which is constrained by link capacities. Those authors assumed that the guaranteed services in the NoC are being preserved at the desired level, and rate allocation of best-effort sources is the main role of the optimization problem. The strategy adopted was to regulate the best-effort source rates with a solution of the optimization problem. It was led to an iterative algorithm that can be used to determine optimal BE source rates and thereby as a means to control the congestion of the NoC. A centralized controller can implement the proposed algorithm.

The technique based on a distributed hardware and software congestion control was proposed by [4]. The proposed system is composed of two NoCs: a data network and a control network. The first one is a network with virtual cut-through switching for application data traffic, and the second one is a specialized network for control and distributed Operating System (OS) services, such as controlling the traffic shaping parameters. These OS services are implemented on small microcontrollers, and included in the network interfaces of the control network. Those authors have introduced Regional Congestion Awareness (RCA), which exploits non-local and local congestion information. The technique adopts a lightweight monitoring network that aggregates and transmits metrics of congestion throughout the network so that each router has a better picture of network hotspots.

Authors [7] proposed a technique based on Ant Colony Optimization (ACO) that was inspired by the related research on the behavior of real-world ant colony. ACO-based adaptive routing has been applied to achieve load balancing with historical information. However, the cost of the ACO network pheromone table is too high, and this overhead grows fast with the scaling of NoC. To fix this problem, those authors proposed a Regional ACO-based routing (RACO) with static and dynamic regional table forming technique to reduce the cost of the table, share pheromone information, and adopt a lookahead model for further load balancing.

Note that all strategies described above are based on some measurement of the network state. The authors [4] proposed a distributed hardware and software congestion control. The authors [1] proposed load balancing in based on link utilization, and [2] suggested a flow control scheme based on source utility maximization. Finally, a Regional ACO-based routing with static and dynamic regional table forming technique was the solution suggested by those authors [7]. We summarize these works in Table I, highlighting the main features of them.

TABLE I. SUMMARY OF TECHNIQUES FOR LOAD BALANCING ON NETWORK-ON-CHIP.

|     | Year | Technique | Strategy |
| --- | --- | --- | --- |
| [4] | 2006 | Distributed hardware and software congestion control. | Composed of two NoCs: data and control. |
| [1] | 2007 | Load balancing based on link utilization. | Control of offered load based on critical shared resource utilization measured. |
| [2] | 2007 | Flow control scheme based on source rate maximization. | Regulates best-effort source rate with the solution of the optimization problem. |
| [7] | 2010 | Technique based on Ant Colony Optimization (ACO) | Network pheromone tables. |

## III. RTSNoC ARCHITECTURE

Similarly to the networks presented in the previous Section, the major guideline behind the design of RTSNoC is prevent traffic congestion of the flows in the networks, and keep fair the access of these flows to the communication channels. Differently from them, however, RTSNoC was conceived for highly dynamic scenarios and therefore strategies based on resource reservation were ruled out in favor of deterministic scheduling. The basic idea is to embody each flit with routing and scheduling information so that routing is performed flit-by-flit based solely on information locally available at each router in a way that preserves the determinism of the worst-case latency for each path.

The additional overhead of carrying routing information along with each flit does not exceed the amount of resources "wasted" on reservation-based networks. For example, an increase of silicon consumption due to the growing number of bits used to address the routers in the network was in average 0.3%; meanwhile the consumption grows in average up to 4.0% when the `Data` field size of the flit changes from 16 bits to 256 bits [18].

## A. Network concept

The adoption of a routing algorithm and an arbiter that allows alternately access to the router output port ensure predictability for all flows in the network without reservation of network resources, and thereby we focused on finding techniques to turn it feasible. To achieve that goal, the following four assumptions were taken into account:

- The routing is done flit-by-flit and made fairly among the flows that are competing for a communication channel;
- Arbiters must grant priority to flits coming from distant routers;
- To minimize the competition for communication channels between routers, up to eight communication channels are available for each router, to explore the sense of locality; and
- Since the routing is done flit-by-flit, the buffers are placed only on the end points, minimizing the side effect of growing on silicon area.

A key element of RTSNoC design is flit-by-flit routing. Since every flit carries along its destination address, arbitration can be implemented locally on each router for each of its output ports. If there are several packets being routed through the same link, the arbiter will alternate access to the corresponding output port so that each flow gets to forward one flit at a time. Hence, the flits from different packets are interleaved in the network. Conversely, circuit-based or wormhole-routing networks would block the output port at least until the end of a packet. Additionally, the flit-level, interleave of flits routing strategy of RTSNoC largely simplifies buffering: routers only have to implement a single-flit buffer for each output port.

## B. Routing and arbiter algorithms

Routing of flits is performed using the XY routing algorithm, which ensures in-order, deadlock-free delivery for 2-D orthogonal networks [19][20]. Since there is only one routing path[1] for the communication between any two cores in the network, the flits from a packet are delivered at the destination in the same order that they have been injected at the origin.

Each output channel of RTSNoC's router has its own "arbiter" to receive and manage the requests generated by the routing controllers at the input channels (Figure 1). The arbiter is similar to the weighted round-robin scheduling algorithm. When the system starts, each channel receives a different priority level. The highest priorities are given to the channels NN, SS, EE, and WW, since they are used to interconnect other routers in a 2-D regular mesh network, as depicted in Figure 2. Any flit has its routing request attended if it has the highest priority, or if there are no other requests in the arbiter. Priority channels NN, SS, EE, and WW might send more than one flit sequentially. The amount of flits that a priority channel may send sequentially is related to the amount of requests that may happen at the same time on these priority channels.

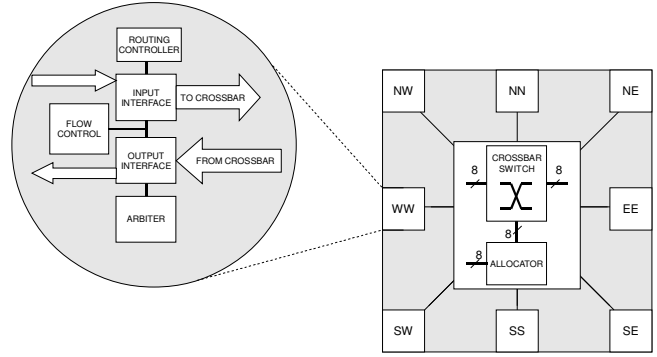

Figure 1. Block diagram showing the internal structure of the router.

Figure 2 shown an example where five cores (gray circles numbers 1, 2, 3, and 4) are sending packets (arrows $\sigma_1$, $\sigma_2$, $\sigma_3$, and $\sigma_4$) to the same destination node (black circle, number 5). Channel EE of router 0 has priority to send up to 2 flits sequentially, and channel SS of router 2 has priority to send up to three flits.

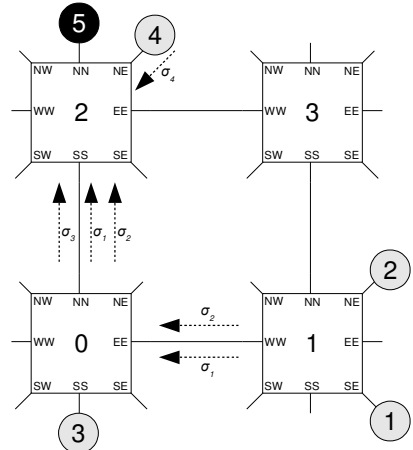

Figure 2. Example of priority channels.

Once the request is attended, the channel that requested the sending of a flit receives the lowest routing priority level and may only send other flits if there is no other flit waiting for routing. The exceptions are the priority channels when used as interconnection between routers. In this case, counters with the priorities are decremented, and the lower priority will grant when they reach value 0.

RTSNoC adopts a 2-D orthogonal topology compatible with the XY routing algorithm. Its routers can be configured at synthesis-time to feature from five to eight interconnection ports. By convention, ports are named after the cardinal points and can be connected either to a core or to another router in order to build larger networks.

We understand that the complexity of some elements in a router grows exponentially with the number of ports, but we had empirical evidence that the "placement" of the cores in the network plays a major role in real applications. We therefore decided to support up to eight ports per router, thus enabling designers to connect cores that will communicate more intensively with each other on the same router, forming clusters as those shown in Figure 3. This design also reduces average the number of hops in the network, which in turn reduces the average communication latency.

---

[1] Flits are always routed first in the X-axis, and subsequently in the Y-axis.

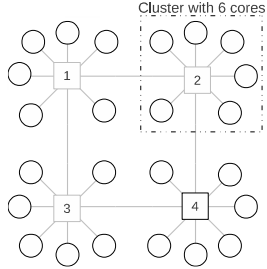

Figure 3. A regular mesh NoC with 4 routers and 24 cores.

## IV. LATENCY ANALYSIS

This Section introduces latency basic concepts with a view to making clear how the interleaving-of-flits method can be suitable to improve the average latency for variable bitrate flows that compete for the same communication channel.

As mentioned earlier, NoCs adopted in real-time platforms must ensure guarantees on latency for individual core-to-core communication in the network. The latency of the network is the time required for a packet to traverse the network, from the time the header of the packet arrives at the input channel to the time the tail of the packet departs the output channel [16]. The latency can be separated into two components:

$$L = T_h + \left(F/b\right) \quad (1)$$

in which $L$ is the packet latency, $T_h$ is the time required for the header of the packet to traverse the network and $(F/b)$ is the time for the packet of length $F$ to cross a channel with bandwidth $b$. In absence of contention, header latency might be seen as the sum of two factors determined by the topology: router delay and number of routers in a path between origin and destination. Based on these two factors, the Eq. (1) can be re-written as follows:

$$L = \left(H_{path} \cdot t_r\right) + \left(F/b\right) \quad (2)$$

in which $H_{path}$ is the number of hops in the path and $t_r$ is the router delay. For simplicity, we do not include in the Eq. (1) and (2) the wire delay across the physical channel, even not the distance from the source and destination of a packet.

Let us suppose that there are three requests to send packets through the same communication channel at instant $t_0$, as shown in Figure 4-a, and the sequence of scheduling for those requests on instant $t_0$ is `packet 1`, `packet 2`, and `packet 3`. We define "Interleave of flits" the method in which the packets from all requests are broken into flits, and these flits have been sending through the channel, one flit from each packet at a time. The interleaving of flits for this example is shown in Figure 4-b and a wormhole switching of those packets is shown in Figure 4-c.

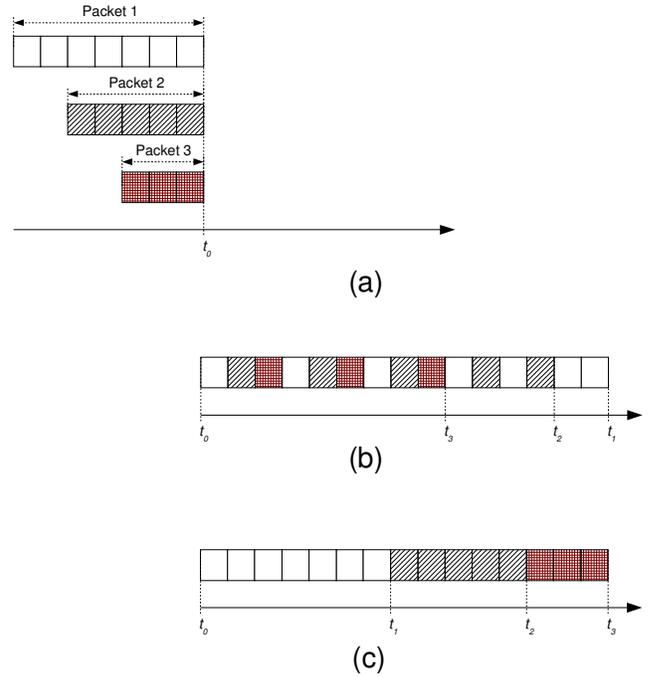

Figure 4. Example of interleave of flits in the same channels.

Note that the time to traverse the channel for the `packet 3` using interleave of flits is ($t_3 - t_0$), which is lower than the time spent in the wormhole switching. It means that the smaller size packets have better average latency when the method of interleave of flits is used, while bigger size packets have their average latency increased.

This is a trade-off when the interleave of flits is adopted. The method is suitable for the systems on which short packets must have their average latency improved, despite the growing in the number of bigger packets that might happen in he network. Recalls that the systems addressed in this paper have flows with variable bitrate, and packets 1, 2, and 3 depicted in the example above eventually have different sizes along the time. These were characteristics we noticed on the industrial R&D projects carried out in the last decade. One example of these projects is presented in Section V-C, on which real-time flows with variable bitrate compete for the same network resources.

As we mentioned in Section I, multimedia applications usually are treated as best-effort flows on NoCs due to silicon consumption issues. For best-effort networks, the communication channels are shared by several flows. We did simulations with the Equation (2) for two hypothetical networks: one network with best-effort support and another one based on interleaving of flits. We have adopted as assumption that the best-effort NoC was implemented with a round-robin arbiter and wormhole switching. The latency for a flow σ$_i$ was calculated according the following equation:

$$L = (H_{path} \cdot t_r) + \frac{F}{|b - b_{occupied}|} \quad (3)$$

in which $|b - b_{occupied}|$ is the bandwidth available for the flow under analysis $\sigma_i$, taking into account that $b_{occupied}$ of the whole bandwidth $b$ has been used by other flows (offered traffic). Remember that for the wormhole switching, if a packet requests a communication channel that is being used by another packet, it must wait for another packet finish the transmission and release the resource communication channel before it starts its transmission.

The expression of latency for a hypothetical network based on the interleave of flits must take into account that there is no resources reservation in the network such happen for wormhole switching, and the flits from a packet may be interleaved with flits from other packets. Hence, the header latency $T_h$ must consider that $N$ packets might compete at each router in the path, from its origin up to its destination ($H_{path}$). It means that, under the latency point of view, the packet size grows $N$ times. Thus, the expression of latency is given as follows:

$$L = (H_{path} \cdot t_r) + N \cdot \left(F/b\right) \quad (4)$$

A simulation of a SoC using these hypothetical networks was done based on the equations (3) and (4). It was done considering the following conditions:
- The number of routers in the path under analysis, for both networks, is 4;
- The router delay is the same for both networks, with the value 3;
- There are three packets requesting the same resources for both networks. One of them, called σᵢ, is the packet under analysis with fixed size of 100 flits, and the other two packets have variable sizes, from 0 up to 64k flits.

Figure 5 depicts the results generated by the simulation with Equation (3) (gray color), and Equation (4) (black color). As expected, the latency for σᵢ was the same, when no other flows request the same resources (offered traffic = 0). With other two flows competing for the same resources, the latency for σᵢ in the NoC based on interleave of flits grows nearly three times; however, it keeps constant up to the maximum bandwidth usage. On the other hand, the latency of σᵢ in the best-effort network has grown when the offered load to the network is nearly 70%, as a result of network congestion for that flow under analysis. The result for interleave network was expected because the latency depends on the number of flits of the flow under analysis, and the number of flows that request the same resources, as shown in Equation (4).

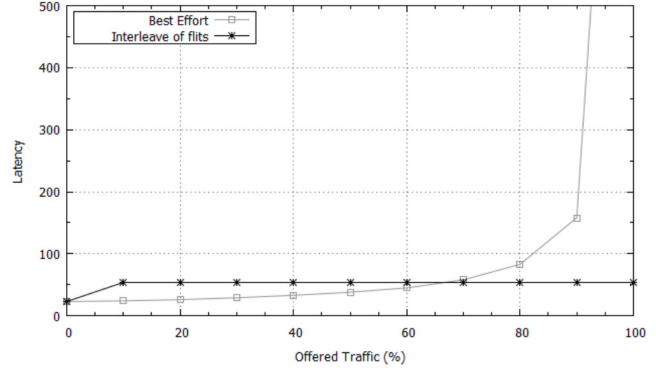

Figure 5. Simulation of interleaving for hypothetical BE network and the network based on interleaving of flits.

Similar result was introduced by [1], as shown Figure 6. The Figure shows network latency of a best-effort connection as a function of offered traffic measured for a single connection in a Æthereal NoC. The graph shows that latency is small and almost constant up to a certain turning point after which the latency grows steeply. This point is nearly at 75% of the offered load, before saturate. In that example, the latency saturates at 2600 $ns$ because queuing between processing units and network interfaces is not taken into account by the authors.

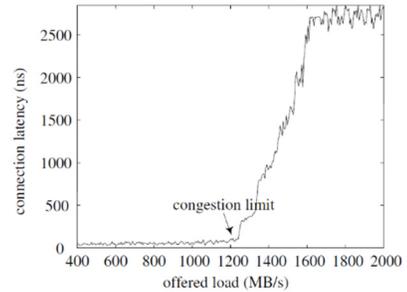

Figure 6. Network latency of an Æthereal BE connection as a function of offered load. Reference: [1].

Based on the information exposed above, we understand that the method of interleave of flits can reduce the average latency for the variable bitrate platform mentioned in Section I. Next Subsection introduces the WCL analytic model of the NoC we proposing in this paper.

### A. WCL analysis

This Subsection introduces the additional latency that a NoC contributes to the execution time of the program instructions. The impact of NoC on the communication latency among the cores is depicted for the networks introduced in Section II. These results are based on the survey published by [8] and used in our paper as a reference. Similar analysis in RTSNoC is introduced, after the analysis over the other NoCs, followed by a benchmark relating these networks.

Essentially, there are four distinguished types of network accesses in terms of the amount of data to be transferred: read/write of single-word transactions or read/write of block transactions. The execution time of a transaction involving a network access includes the time spent traversing the network ($T_{noc}$) and the time spent accessing the remote core ($T_{core}$):

$$T_{transaction} = T_{NoC} + T_{core} \quad (5)$$

($T_{core}$) depends on the characteristics of the cores connected in the network. A transaction through the network may include up to three different delays. One delay is related to the time waiting before getting access to the network ($T_{wait;req}$). Another delay is related to the transaction request sent through the network ($T_{req}$). Finally, a reply should be send back, depending on the case, that would also require some waiting time for gaining access to the network ($T_{wait;reply}$) and some time to transfer the reply through network ($T_{reply}$), back to the requesting node. In general, the contribution of the network to the latency of a transaction may be given by:

$$T_{NoC} = T_{wait;req} + T_{req} + T_{wait;reply} + T_{reply} \quad (6)$$

Equation (6) is a general expression and may be adapted according the type of transaction in the network.

### B. WCL analysis on RTSNoC

The worst-case latency for a packet that belongs to a flow $\sigma_i$ in the RTSNoC network is defined as the sum of the latency experienced by all flits that belong to the same packet, on the path between an origin node and destination node with $h$ routers. Our analysis for the WCL for packets in RTSNoC is based on the Equation (1) introduced in Section IV. The packet latency can be separated into two components: the time required for the header of the packet to traverse the network and the time for a packet of length $F$ to cross the channel with bandwidth $b$.

The first flit of a packet in the RTSNoC is the header flit, and it may has a different latency than other flits of the same packet. It happens because once the first flit reach the destination node, than the other flits subsequent to it will be routing with the priorities established by the first flit in the path. Thereby, if there are no changes on other flows, then the payload and tail flits will have the same latency, that might be different from header flit.

The first latency analyzed in this Section is related to header flit. First, let us use as reference the bandwidth $b$ with one flit per clock cycle. Second, the latency to forward a flit from an input channel to an output channel is two clock cycles in the RTSNoC router [18]. If $N$ flows are competing for the same output channel in a router, then one of their flits is granted at each arbitration cycle. Hence, the maximum latency expected for the header flit, $L_{header}$ that belongs to a packet from the flow $\sigma_i$ is given by the following expression:

$$L_{header} = \sum_{i=1}^{H_{path}} (N_i \cdot t_r) = \sum_{i=1}^{H_{path}} 2N_i \quad (7)$$

Let's take into account the following assumptions: (*i*) the payload flits will be routing with the same priorities established by the header flit on routers in the path between the origin node and destination node, and (*ii*) all of the flows than might compete for the same resources are sending their packets to the same destination. Hence, the latency of payload and tail flit is given as following:

$$L_{payload;tail} = \frac{k(f-1)}{1/2} = 2k(f-1) \quad (8)$$

in which $k$ is the number of packets from other nodes in the whole network that are competing for the same destination node in the network and $f$ is the amount of flits of the analyzed packet. From the expressions (7) and (8) is possible to find out the worst-case latency for any packet in the RTSNoC network, as following:

$$W_{packet} = \sum_{i=1}^{H_{path}} 2N_i + 2k(f-1) + 2B \quad (9)$$

in which $B$ is the buffer size at network interfaces (FIFO memories). The buffer size was multiplied by two because we are considering that is possible that both memories might be not empty with other flits, even in the origin node interface as in the destination node interface.

Note that the parameters in Eq. (9) are well known. Recall that the XY algorithm, implemented in the routers, is a static algorithm and imposes all flits that belong to the same packet must be routing by a unique path. Due to this algorithm's pattern, the maximum value of $N$ will always be the same because the number of cores and routers in the path are fixed. The parameter $k$ is also well known due to the size of the network, and hence is possible to presume the maximum number of flows from other nodes in the whole network that might compete for the same destination node. Furthermore, the parameter $B$ is defined at compilation time and the parameter $f$ is know by origin node. It means that the hard real-time flows designed considering the absolute WCL of RTSNoC will always meet the requirements of the associated hard real-time tasks, so no deadline can be lost due to network contention.

### V. EXPERIMENTAL RESULTS

#### A. Evaluation of WCL and throughput

This Section introduces the results from experiments done with a RTSNoC network composed by four routers and synthesized in a FPGA device from Xilinx manufacturer (XC6VLX75T-3-FF484). The objective was to perform measurements of latency and throughput, in order to evaluate the results with the expected theoretical values to this network. The latency measurement of a packet was done considering the number of clock cycles since the header flit is injected in the input channel at origin node, up to the arrival of the tail flit a the output channel at the destination node. The network has four routers, called 0, 1, 2, and 3, depicted by white squares in Figure 7. Twenty-four cores were connected in the network and depicted by circles (0 up to 23). Five of them generate packets to the same destination node. These five cores, called 3, 7, 13, 18, and 23, are gray circles in Figure 7, whereas the destination node is the black circle called 12.

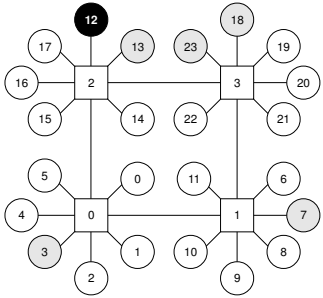

Figure 7. Experimental network with 4 routers and 24 cores.

The packets generated by those five cores have six flits. Table II shown the flits generated by the cores, where the header and tail flits have most significant byte as $4_h$, whereas payload flits have value $0_h$. Furthermore, two less significant bytes of the flits that belong to a packet have values to distinguish them from other flits. For example, flits that belong to a packet from flow $\sigma_i$; they have less significant bytes $3X_h$, in which X means the number of the flit, such as 1, 2, etc.

TABLE II. PACKETS ADOPTED TO EVALUATE THE WCL AND THROUGHPUT.

| Flit type | Packets per flow | | | | |
|---|---|---|---|---|---|
| | $\sigma_3$ | $\sigma_7$ | $\sigma_{13}$ | $\sigma_{18}$ | $\sigma_{23}$ |
| Header | $40831_h$ | $40871_h$ | $408D1_h$ | $408E1_h$ | $408F1_h$ |
| Payload | $00832_h$ | $00872_h$ | $008D2_h$ | $008E2_h$ | $008F2_h$ |
| | $00833_h$ | $00873_h$ | $008D3_h$ | $008E3_h$ | $008F3_h$ |
| | $00834_h$ | $00874_h$ | $008D4_h$ | $008E4_h$ | $008F4_h$ |
| | $00835_h$ | $00875_h$ | $008D5_h$ | $008E5_h$ | $008F5_h$ |
| Tail | $40836_h$ | $40876_h$ | $408D6_h$ | $408E6_h$ | $408F6_h$ |

The latency and throughput measurement were done on flit generated by core 7, according two assumptions: (*i*) cores 3, 13, 18, and 23 generate packets all the time, and (*ii*) core 7 generates only one packet on which the latency and throughput measurements are done. Figure 8 depicts the results from the simulation tool ISim, from Xilinx manufacturer. The experiments were done with a clock frequency of 100MHz, such that clock cycles have 10 *ns*. The header was injected in the input channel of router 1 on the falling edge of clock (i_CLK) as shown in Figure 8 by a letter A and a dashed square. The header flit was delivered after twelve falling edges of clock at the output channel of router 2, where the core 12 was connected (letter B and a dashed square).

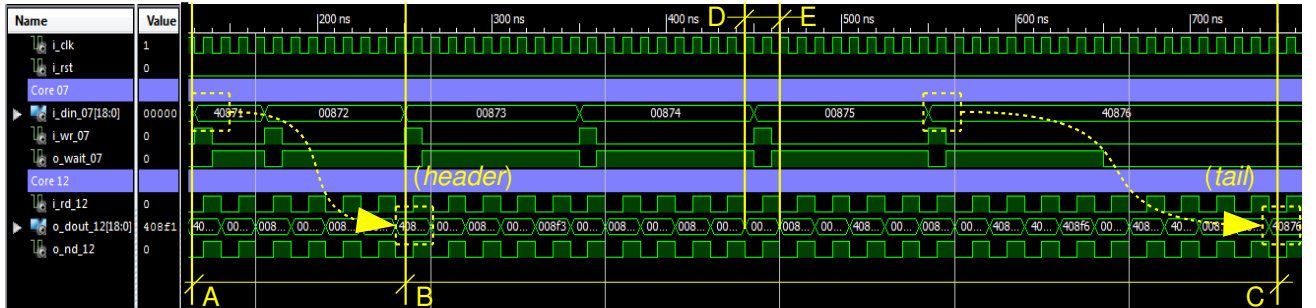

Figure 8. Experimental network with 4 routers and 24 cores.

RTSNoC adopts XY routing, and hence all flits generated by the core 7 follow the path across routers 1 → 0 → 2. The header flit took two clock cycles to be routed in router 1 because there were not competition for the same output channel on that router. The flit took four clock cycles to be routed in router 1, due to the competition with flits from core 3. Finally, the header flit took six clock cycles to be delivered on output channel where core 12 were connected, due to the competition with flits from cores 13 and 3 (also received in channel SS of that router). Remember that this is the theoretical worst-case latency. We did several simulations and measurements with different latency values. However, the latency measured for header flit depicted in Figure 8 was the worst-case.

Equation (9) allows us to find out the theoretical value of latency for the packet shown in Figure 8. We are considering that the reading process of flits is done as soon as the flits are delivered to the destination node and without buffer overflow in the network interfaces. Thereby, the latency contribution of 2B is zero in Equation (9) and the theoretical value of WCL is:

$$WCL_{07} = 12 + 2.5.(6 - 1) = 62 \qquad (10)$$

Note that the latency to deliver the packet was sixty-two clock cycles: twelve for the header flit and fifty for the other flits. In this experiment, we found out the timing to reach the theoretical worst-case latency; other experiments, as expected, gave us lower values and never bigger than WCL. Furthermore, the flits were delivered in the same sequence that they were injected in the origin node. It was expected since the XY algorithm, implemented in the routers, is a static algorithm and imposes all flits that belong to the same packet to be routing by a unique path. Figure 8 also shown that the interval between `D` and `E`, when a flit is delivered, has two clock cycles, and this is the throughput mentioned earlier.

### B. Latency evaluation

This Subsection introduces two simulations that were done using the expression of latency for RTSNoC given in Equation (9) and the expected latency for BE flows in the Æthereal network. Two factors guide us to choose Æthereal network as reference to this evaluation. First, it is one of most cited networks on scientific papers related to NoC[2]. Second, it offers best-effort services allowing the flows treated as BE to use reserved, but unused, time slots time slots to improve the throughput of these flows.

The latency for the Æthereal network was calculated using the Equation 3, in which the bandwidth was based on the throughput expression published in the Survey written by [8]:

$$Throughput_{AEthereal} = \frac{(p.n)}{P.s} \quad (11)$$

For Equation (11), $p$ is the number of time slots assigned to a virtual circuit, n is the number of packets in a transaction, $P$ is the period of time slots in a time schedule and $s$ is the time slot duration in clock cycles. In simulations, we have adopted these parameter with values shown in Table III, and both networks have twenty-three cores. We have considered that Æthereal network support one processing unit per router, and hence, the maximum path for it is eight hops ($5 \times 5$ mesh network); meanwhile RTSNoC with twenty-three processing units has four routers and two hops as maximum path ($2 \times 2$ mesh network). For the simulations we have considered the following assumptions:

- There are three critical flows in the Æthereal network;
- There are four time slots in the network, and three of them were allocated for each critical flow;
- Two simulations were done for Æthereal network. In the first simulation, we are considering that each critical flow uses 60% of its time slot, and the BE flows can use the remaining 40%. For the second one, we are considering that BE flows can use only the time slot reserved for them;
- There are three packets requesting the same resources for both networks. One of them, called $\sigma i$, is the packet under analysis with fixed size of 9 flits, and the other two packets have variable sizes, from 0 up to 64k flits; and
- A single time slot from Æthereal has the same bandwidth than a RTSNoC channel.

The assumptions above were chosen because they are quite close to the real cases we have been investigating, such the case study that will be introduced in Subsection V-C.

TABLE III. PARAMETER LIST ADOPTED FOR THE LATENCY SIMULATIONS CONSIDERING EIGHT HOPS PER NETWORK.

| Parameter | Value |
|---|---|
| n | 1 |
| f | 9 |
| p | 1 |
| P | 4 |
| s | 1 |

Figure 9 depicts the simulation results. Black curves are related to Æthereal simulations, and the gray ones are related to minimum and maximum latency expected for the flow $\sigma i$ in the RTSNoC network. Note that Æthereal network with allocation of unused GS time slots has best performance, in terms of latency and congestion limit than the another one that uses only the BE time slot. RTSNoC has better load balancing for offered traffic over than 70%, when compared with an Æthereal network with exclusive best effort time slots. Meanwhile, the load balancing on RTSNoC is better for a offered traffic over than 80% when compared with Æthereal that support BE flows in unused GS time slots.

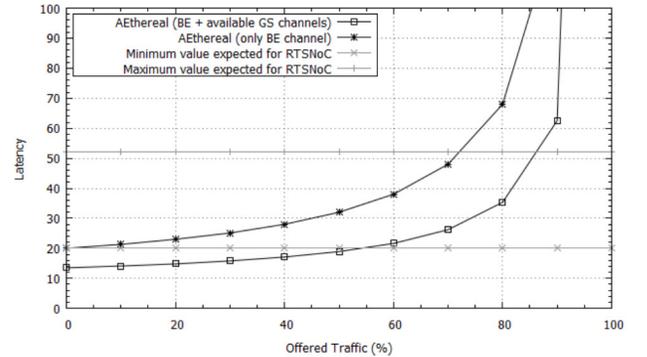

Figure 9. Expected latency *vs.* offered traffic for Æthereal and RTSNoC networks.

The results depicted in Figure 9 allow us to argue that RTSNOC achieved its goal offering better load balancing for high traffic in the network. It means that RTSNoC is suitable for full bandwidth usage; otherwise, other solutions might have lower latency.

### C. Case study - PABX

We experimentally evaluate RTSNoC by deploying an industrial application of a Private Automatic Branch eXchange (PABX), which we call the PABX SoC. This equipment is an example of a system with variable bitrate that, usually, works over 60% of its maximum load in the communication channels among the cores that compose it. According the manufacturer, if the PABX equipment is under a high traffic of phone call procedures, the deadline loss of flows might affect the overall quality of the PABX system. Furthermore, the quality perception might be an important issue for some users.

---

[2] Information available at www.ieeexplore.com.

*1) PABX SoC structure:*

The PABX SoC targets FPGA-based telecommunication systems with the following necessary features to implement the digital PABX:
- One single tone Hz generator;
- One Dual-Tone Multi-Frequency (DTMF) generator;
- One conference switch;
- One DTMF detector;
- One interface to E1 link;
- One subscriber interface with support for thirty-three subscribers;
- One VoIP (acronym of Voice over Internet Protocol) interface;
- One single tone detector; and
- One Time-Division Multiplexing (TDM) bus synchronization core.

The original implementation of this SoC is shown in Figure 10-a. The traditional TDM communication was replaced by the RTSNoC in this SoC, as shown in Figure 10-b. Other equipment, such as E1 call generator, telephone equipment and external hardware for subscriber were used in the experiment but we will not detail them in this document.

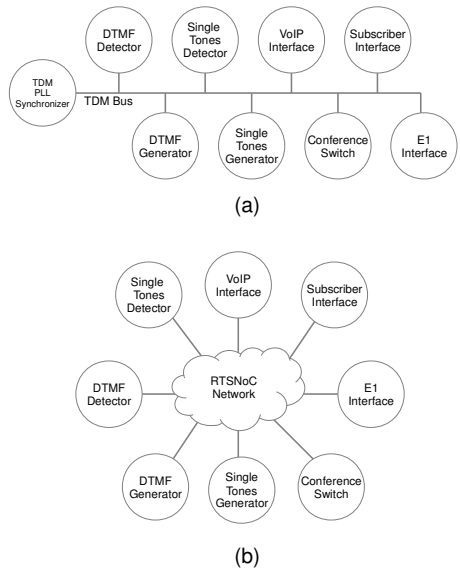

Figure 10. PABX structure: (a) block diagram showing the original structure of PABX SoC based on TDM; and (b) block diagram showing the PABX SoC using RTSNoC as interconnection for the cores.

*2) PABX implementation and results*

We synthesize the PABX SoC using Xilinx's ISE version 13.1 targeting a Virtex 6 FPGA. To analyze the maximum operating load, we generate thirty telephone calls on E1 link, and transmit it to the Subscribers interface. The incoming calls are randomly generated with short durations. For each incoming call, the PABX cores exchange messages related to call identification, DTMF signal generation and detection, and call forwarding further the voice. Based on the characteristics of the PABX SoC, we compute the WCL of the packets, which results in a worst-case latency of 420 *ns*.

We measure the latency across the network on the PABX SoC received shows the latency variation of synchronization packets received at the single tone detector, which placement was done at NW port of an RTSNoC router (Figure 11). It was chosen due to its simplicity to check possible distortions related to synchronization issues. Our measurements showed the latency variations of packet transmission from 288 – 317 *ns*, which is less than the WCL of 420 *ns*, i.e. the average latency was 28% lower than the WCL. Furthermore, the WCL rarely was achieved, and never was bigger than 420 *ns*.

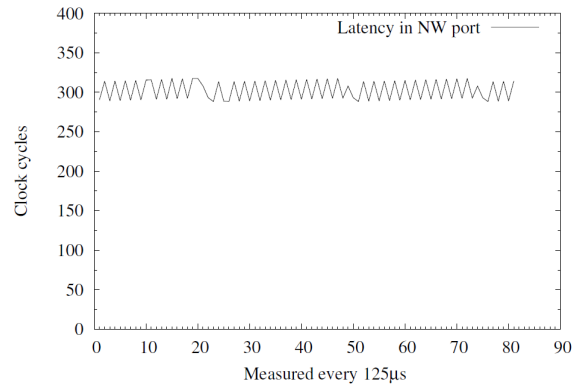

Figure 11. Latency of synchronization packets at port NW.

## VI. CONCLUSION

This paper presented the design and evaluation of a Network-on-Chip that offers load balancing for SoCs dedicated for applications that require variable bitrate communication. The design was based on a connectionless strategy on which flits from different communication flows are interleaved in the same communication channel. Each flit carries routing information that is used by routers to perform arbitration and scheduling of the corresponding output ports to balance channel utilization.

Despite the growing on silicon consumption caused by the adopted strategy, experiment's result demonstrates in Subsection V-B that the average latency is kept lower the WCL boundary when the offered traffic is higher than 80%, what does not happen on regular BE schemes. We also analytically demonstrate in Subsection IV-B that real-time flows designed considering the absolute WCL of RTSNOC will always meet the requirements of flows associated with real-time tasks so that no deadline can be lost due to network contention.

Based on the results introduced in this paper, we understand that RTSNoC is a suitable solution for SoCs based on Network-on-Chip that demand for load balancing when the network is under high traffic of variable bitrate flows. For low traffic of variable bitrate, a regular BE scheme can be better in terms of average latency. The experiments also have validated our hypothesis, mentioned in Section I, that the load balancing could be improved at design-time for these SoCs with high traffic of variable bitrate flows, and without check the network state at running time.